\documentclass[conference, comsoc]{IEEEtran}
\usepackage{cite}
\usepackage{amsmath,amssymb,amsfonts,utfsym}
\usepackage{graphicx}
\usepackage{textcomp}
\usepackage{xcolor}
\usepackage[inline, shortlabels]{enumitem}
\usepackage{booktabs}
\usepackage{multirow}
\usepackage{rotating}
\usepackage{float}
\usepackage{subfig}
\usepackage[hidelinks]{hyperref}
\usepackage{xurl}
\usepackage{algorithm}

\usepackage{algorithmic}
\newcommand{\KwPubIn}{\textbf{Public Inputs: }}
\newcommand{\KwPrivIn}{\textbf{Private Inputs: }}
\newcommand{\KwAssert}{\textbf{assert }}

\usepackage{pifont}

\usepackage{caption}
\usepackage[inline, shortlabels]{enumitem}

\def\BibTeX{{\rm B\kern-.05em{\sc i\kern-.025em b}\kern-.08em
    T\kern-.1667em\lower.7ex\hbox{E}\kern-.125emX}}

\renewcommand{\baselinestretch}{0.995}

\graphicspath{ {./figures/} }

\begin{document}

\author{
\IEEEauthorblockN{
 Panagiotis Michalopoulos\IEEEauthorrefmark{1},
 Anthony Mack\IEEEauthorrefmark{1},
 Cameron Clark\IEEEauthorrefmark{1},
 Linus Chen\IEEEauthorrefmark{1},
 Johannes Sedlmeir\IEEEauthorrefmark{2},
 Andreas Veneris\IEEEauthorrefmark{1}\IEEEauthorrefmark{3}
 }

 \IEEEauthorblockA{\IEEEauthorrefmark{1} Department of Electrical and Computer Engineering, University of Toronto, Toronto, Canada \\ \{p.michalopoulos, anthony.mack, camo.clark, linus.chen\}@mail.utoronto.ca, veneris@eecg.toronto.edu}
  \IEEEauthorblockA{\IEEEauthorrefmark{3} Department of Computer Science, University of Toronto}
 \IEEEauthorblockA{\IEEEauthorrefmark{2} Department of  Information Systems, University of Münster, Münster, Germany \\ johannes.sedlmeir@uni-muenster.de}
}

\title{Balancing Compliance and Privacy in Offline CBDC\\Transactions Using a Secure Element-based System}

\maketitle

\begin{abstract}
Blockchain technology has spawned a vast ecosystem of digital currencies with Central Bank Digital Currencies (CBDCs) -- digital forms of fiat currency -- being one of them. An important feature of digital currencies is facilitating transactions without network connectivity, which can enhance the scalability of cryptocurrencies and the privacy of CBDC users. However, in the case of CBDCs, this characteristic also introduces new regulatory challenges, particularly when it comes to applying established Anti-Money Laundering and Countering the Financing of Terrorism (AML/CFT) frameworks. 
This paper introduces a prototype for offline digital currency payments, equally applicable to cryptocurrencies and CBDCs, that leverages Secure Elements and digital credentials to address the tension of offline payment support with regulatory compliance.
Performance evaluation results suggest that the prototype can be flexibly adapted to different regulatory environments, with a transaction latency comparable to real-life commercial payment systems. Furthermore, we conceptualize how the integration of Zero-Knowledge Proofs into our design could accommodate various tiers of enhanced privacy protection.
\end{abstract}


\section{Introduction}
The emergence of blockchain has catalyzed the  digitalization of payment services~\cite{Adrian2019}  through the creation of cryptocurrencies, such as
stablecoins~\cite{exec_order}, representing a significant shift in the global financial landscape.
This shift, primarily led by private entities, has created concerns among central banks over private control of payment infrastructures, financial stability, and monetary  sovereignty~\cite{Adrian2019,bis2023a}.
To maintain the effectiveness of their tools in providing financial stability, central banks are heeding the idea of protecting their raison d'être: more than 90\,\%~\cite{acouncil_tracker} of central banks are actively investigating digital versions of fiat money, known as {\em Central Bank Digital Currencies (CBDCs)}~\cite{bis2023a} and the potential of incorporating blockchain and distributed ledger technologies in their designs~\cite{bis2024,wef2019}.
Such digital instruments would enhance monetary policy and payment efficiency, safeguard monetary sovereignty, and improve financial inclusion~\cite{Allen2020,bis2020,Bindseil2020}.

An important feature for both cryptocurrencies and CBDCs is the ability to perform value transfers between parties without requiring a connection to an online ledger or other network infrastructure~\cite{bis2023b,Androulaki2024,Beer2024,Bean2024,Christodorescu2020,Yang2022,Jie2024}. For the former, this functionality can assist in addressing the scalability problem of blockchains~\cite{Jie2024}, while for the latter it ensures payment accessibility during network outages or in regions with limited connectivity, inclusion of under-banked populations, and enhances system resilience during disasters or infrastructure failures~\cite{bis2020,bis2023b}. 

However, {\em offline CBDCs} also introduce unique regulatory challenges. Their potential to offer transactions without any connectivity ({\em i.e.,} communication via a third party authority) gives them a level of privacy comparable to that of physical cash.  As such, ensuring compliance with Anti-Money Laundering and Combating the Financing of Terrorism (AML/CFT) regulations becomes a challenge, creating an apparent {\em trade-off} between privacy and transparency.
For instance, real-time transaction screening and monitoring that could uncover illicit activities are not straightforward procedures in such offline scenarios~\cite{Michalopoulos2025}. Additionally, the lack of identification of transacting partners in cash payments today complicates AML/CFT compliance as it effectively anonymizes counterparties and impedes the traceability of monetary flows.   
Therefore, the question of how digital identity mechanisms can be integrated into offline CBDC systems in a way that enables compliance while preserving user privacy arises as an additional challenge. 

This paper addresses these intertwined challenges by arguing for a \emph{compliant-by-design} hardware/software platform~\cite{Michalopoulos2025,pocher2022}. This scheme aims to create inherently compliant payment instruments with digital identity mechanisms. To that end, we introduce an open-source\footnote{https://github.com/Veneris-Group/offline-cbdc-prototype} implementation of an offline CBDC prototype that leverages secure hardware and Zero-Knowledge Proofs (ZKPs) to offer regulatory-compliant and privacy-preserving offline transactions.
Secure hardware is used to correctly execute an offline payment protocol and to enforce regulatory measures (\emph{e.g.,} holding limits and transaction tracking). Meanwhile, ZKPs allow for privacy-preserving identity verification and compliance checks, such as conditional payments based on user attributes stored within and outside the Secure Element (SE) -- \emph{e.g.,} age-restricted purchases. The proposed system remains relevant to cryptocurrencies, since similar compliance requirements could prove equally important for cryptocurrencies and decentralized finance~\cite{Duffie2025, Xiong2025}.

In more detail, the paper's contributions include:
\begin{itemize}
    \item an offline CBDC prototype implementation demonstrating how monetary functions can be combined with digital identity attestations to enforce regulatory constraints. Specifically, the requirements, components, and operations of the prototype are presented;
    \item a performance evaluation of the prototype on real-world, resource-constrained devices that demonstrates that transaction latency remains below 5 seconds, \emph{i.e.,} comparable to modern payment systems;
    \item a conceptual discussion on how the integration of ZKPs can further elevate privacy benefits while maintaining regulatory compliance; and
    \item a security analysis that derives the payment integrity properties of the system but also delves into the privacy/transparency trade-off.
    
\end{itemize}

The remainder of this paper is structured as follows: Sec.~\ref{sec:background} presents background on CBDCs and AML/CFT compliance, along with related work. Sec.~\ref{sec:overview} gives an overview of the hardware and software components of the proposed solution and Sec.~\ref{sec:details} presents the design details of our prototype. Sec.~\ref{sec:eval} evaluates the performance of the system, while Sec.~\ref{sec:zkps} discusses the integration of ZKPs. Sec.~\ref{sec:sec_analysis} performs a comprehensive security analysis of the system and we conclude with a discussion of avenues for future work in Sec.~\ref{sec:conclusion}.

\section{Background and Related Work}\label{sec:background}	

\subsection{Central Bank Digital Currencies}
CBDCs are the equivalent of ``digital fiat money'' as -- unlike ``commercial bank money'' -- they represent a liability of the central bank. As such, they serve as an alternative form of base money, next to physical cash and reserve accounts of the private sector maintained at the central bank~\cite{Allen2020,Barontini2019,Fanti2022}.
CBDCs can be \emph{wholesale} or \emph{retail}. Wholesale CBDCs are designed for financial institutions (FIs), such as commercial banks or intermediaries (\emph{e.g.,} payment service providers), whereas retail CBDCs are intended for public use.
Retail CBDCs can be administered (\emph{e.g.,} account opening) directly by the central bank, giving rise to a \emph{one-tier} model, or administration can be delegated to FIs in a \emph{two-tier} model~\cite{Allen2020} -- the latter resembling the financial practice today. 
Another common -- yet contested~\cite{Michalopoulos2025} -- classification distinguishes CBDCs between \emph{token-based} and \emph{account-based} according to how users gain access to the CBDC~\cite{Auer2020,Carstens2021}. Token-based systems function through the exchange of verifiable cryptographic tokens with a predefined value and may optionally involve custodians who hold tokens on behalf of users. Account-based systems typically depend on some form of identity verification, such as Know-Your-Customer (KYC) procedures, and make use of account balances. As offline functionality usually applies to retail CBDCs, we focus solely on them in this paper.

\subsection{Offline CBDC transactions}
Offline transactions are payments made in the absence of a connection to an online ledger~\cite{bis2023b}.
The Bank for International Settlements classifies~\cite{bis2023b,Michalopoulos2025} offline CBDCs as \emph{fully offline}, \emph{intermittently offline}, or \emph{staged offline} based on \emph{(i)} whether transaction settlement occurs offline or online, and \emph{(ii)} the duration a wallet can remain offline before synchronization with the online ledger is needed. Fully and intermittently offline systems settle offline and received funds can be immediately re-spent; the former allow unlimited offline duration, while the latter require recurring synchronizations. Staged offline settle online during synchronization. The system presented in this paper falls under the \emph{intermittently offline} category.

Current offline CBDC proposals~\cite{Androulaki2024,Beer2024,Bean2024,Christodorescu2020,Yang2022} rely on secure hardware~\cite{Christodorescu2020,Yang2022}  or a combination of secure hardware and cryptographic primitives~\cite{Androulaki2024,Beer2024,Bean2024}.
In~\cite{Christodorescu2020}, a solution based on Trusted Execution Environments (TEEs) is developed, while in~\cite{Yang2022} a combination of a Secure Element (SE) and a TEE is used. Compared to those approaches, which do not touch upon compliance, this paper focuses on a solution that enables regulatory compliance in a privacy-preserving way.
Further,~\cite{Yang2022} simulates the SE through a micro-controller, whereas we implement our solution on an actual SE.
Compared to~\cite{Bean2024}, which uses custom hardware and physical unclonable functions, we focus on commercial-off-the-shelf smart cards to demonstrate the feasibility of our system on hardware used by the existing payments infrastructure.
The works closest to ours are~\cite{Androulaki2024} and~\cite{Beer2024}. 
In~\cite{Beer2024}, ZKPs are used to provide a private and compliant offline CBDC system that identifies double-spenders in case of an SE compromise. 
In contrast, while acknowledging the potential vulnerabilities of SEs, this paper focuses on the implementation details pertaining to the SE and provides a prototype that could be used as a testbed for further explorations. It also approaches regulatory compliance through the integration of digital identity attestations -- which may be stored external to the SE or in another sandboxed applet --  to the CBDC system. 
In~\cite{Androulaki2024}, a system based on an SE combined with a cryptographic protocol to enable private offline token-based CBDC transactions and double-spender identification is presented. In comparison, we put an emphasis on addressing the trade-off between compliance and privacy through an account-based approach. 
Finally, when compared to~\cite{Androulaki2024,Beer2024}, our proposed system exhibits a constant payment latency that does not increase with transaction or token history, and regulatory flexibility since the same firmware adapts to different AML/CFT risk-tiers.

\subsection{AML/CFT in payment systems}
AML/CFT regulations aim to safeguard the financial system by preventing criminals from hiding the origins of illegal funds through the assignment of responsibilities to regulated entities (\emph{e.g.,} FIs)~\cite{schlatt2022}.
The paper adopts a jurisdiction-agnostic approach on AML/CFT based on the Recommendations of the Financial Action Task Force (FATF)~\cite{FATF}.
Central to FATF's recommendation is the {\em Travel Rule} that requires FIs to share sender and receiver information when transferring funds over a  threshold. 
For offline CBDCs, AML/CFT responsibilities assigned to FIs can include, among others,~\cite{Michalopoulos2025}: \emph{(i)} KYC by identifying clients and verifying their identity; \emph{(ii)} the association of transactions to user identities; \emph{(iii)} the enforcement of offline usage limits (\emph{e.g.,} thresholds on transaction amount, turnover, or balance); \emph{(iv)} the storing of offline transactions; and \emph{(v)} the automatic or manual monitoring of offline transactions (\emph{e.g.,} transaction tracking, graph analysis).

\section{Solution Overview}\label{sec:overview}
    The proposed system enables users to transact offline in a compliant way: {\em value exchange} and {\em transaction settlement} ({\em i.e.,} funds are available for re-spending) occur without the need for the user(s) to connect to the online ledger. Following are the system's main entities and operations.

    \subsection{Ecosystem Entities}
    \subsubsection{Central bank and FIs}
    Without loss of generality, we assume a two-tiered CBDC system. The central bank is the issuer of the CBDC and (in some capacity) oversees the operation of the system. FIs maintain reserve accounts with the central bank and interact with the users of the CBDC. Specifically, they perform the necessary KYC procedures and hold users' accounts.
    
    \subsubsection{Users}
    System participants maintain an account with an FI, possess a \emph{secure device} equipped with an SE, and install a \emph{wallet application} on their mobile phone. The SE can be embedded in their phone or be separate in the form of a smart card that communicates with the wallet application through the NFC interface of the phone. Without loss of generality, in this paper we assume the latter.
    Currently, only some smartphones incorporate embedded SEs, and even for these, programmability remains restricted. Although smartphone manufacturers like Google and Apple are beginning to provide limited API access to their SEs~\cite{apple_se,google_se}, which may allow for smartphone implementations, choosing a smart card allows lowering the barrier to entry to the offline system for users who do not own a niche smartphone or with no smartphone at all.
    Notably, in a recent user study by the Bank of the Netherlands most of the participants (42\,\%) preferred a smart card for offline CBDC transactions~\cite{BoN2025} over the other options.
    
    Since SEs are passive elements, \emph{i.e.,} they do not have their own power supply, the phone combined with the wallet application functions as the \emph{user terminal} for the secure device. It provides power, handles the user interface and communication with the secure element, and facilitates the communication between the secure devices of the two users. In the following, we will use the terms wallet application and user terminal, as well as SE and secure device, interchangeably.
    Since the wallet application is running outside of a protected environment, we consider it to be compromisable. 

    \begin{table}
    \centering
    \renewcommand{\arraystretch}{1.1}
    \caption{Invocation permissions for the SE's operations}
    \begin{tabular}{lccc}
    \toprule
    \bfseries Operation & \bfseries FI terminal & \bfseries User terminal & \bfseries Secure device \\ \midrule
    \texttt{<Withdraw>} & $\checkmark$ &  &  \\
    \texttt{<Request>} &  & $\checkmark$ &  \\
    \texttt{<Accept>} &  & $\checkmark$ &  \\
    \texttt{<Transfer>} &  &  & $\checkmark$  \\
    \texttt{<Receive>} &  &  & $\checkmark$  \\
    \texttt{<Retransmit>} &  &  & $\checkmark$  \\
    \texttt{<Synchronize>} & $\checkmark$ &  &  \\
    \texttt{<Deposit>} & $\checkmark$ &  &  \\
    \bottomrule
    \end{tabular}
    \label{tab:api_perms}
    \end{table}
    
    \subsubsection{Public key infrastructure}
    Every device is assigned a \emph{participation certificate}, which authorizes it to operate inside the system and assigns its role (\emph{i.e.,} either a secure device, a user terminal, or an FI terminal). Roles restrict access to the secure device's API as depicted in Table~\ref{tab:api_perms}. For instance, only an FI terminal can initiate the withdrawal protocol. For simplicity, we assume a single {\em certificate authority} (CA) -- \emph{e.g.,} the central bank -- signs all certificates and that the cryptographic keypair of each certificate is unique per SE to ensure that compromise of a single key cannot help impersonate any other secure device, and to enable enhanced accountability measures.
    
    \subsection{Operations}
    Consider users Alice (``A'', the {\em receiver}) and Bob (``B'', the {\em sender}) who wish to transact offline. If $i\in\{A,B\}$ is the user, let 
    $d_{i}$ denote the user's secure device (\emph{i.e.,} the smart card), and $w_{i}$  the wallet application installed on their phone. 

    \subsubsection{Setup} During initial setup, user $i$ undergoes a one-time KYC procedure with a designated
    institution (such as a government agency or an FI) where $d_i$ generates and securely stores a keypair $(sk_{i},\, pk_{i})$ -- with $sk_{i}$ being the secret key and $pk_{i}$ the public key. 
    After successful KYC verification of the user, the designated institution requests the certificate authority to create and sign the device's participation certificate $c_{i}$ -- thus, certifying that $pk_{i}$ belongs to an authorized secure device. The certificate is stored in $d_{i}$ along with the authority's public key, and the institution stores $(\mathrm{id}_{i},\, pk_{i})$, where $\mathrm {id}_{i}$ is user $i$'s government-issued identifier provided during KYC verification.
    This data could be encrypted for enhanced user protection and decrypted only under a court order.
    Finally, the wallet application is also assigned its own separate certificate (created and signed by the certificate authority) which is used for authenticating itself to the secure device and other wallets.
    
    \subsubsection{Withdraw \& deposit} Users can withdraw funds from their account to their secure device or deposit funds from their secure device to their account. The FI terminal sends the appropriate signed command to the secure device, which in turn verifies it, checks if regulatory constraints are satisfied ({\em e.g.,}  balance/transaction limits), and modifies the balance accordingly. As described later in this paper, the withdrawal protocol can be modified with ZKPs to accommodate private withdrawals, where a user can exchange physical cash for CBDCs at an FI terminal instead of withdrawing funds from their account, similarly to previous CBDC proposals~\cite{Gross2021}.

    \subsubsection{Offline payment} 
    The offline payment protocol between Alice and Bob is divided into two stages: {\em payment initiation} and {\em value exchange}. During the first stage, the two parties use their wallet applications to agree upon the payment amount and roles (\emph{i.e.}, who is the sender and who is the receiver) and authorize the transaction. In parallel, their secure devices check if the transaction can be fulfilled (\emph{e.g.,} sufficient balance) and enforce any regulatory constraints. During the second stage, the funds move between the devices through appropriate balance updates. The protocol employs a mutual authentication subroutine to establish an encrypted channel to ensure the confidentiality and integrity of the exchanged information, as well as to ensure the legitimacy of the transacting parties. Similarly, to address potential communication interruptions that may lead to transaction failures and inconsistent state, the system includes a retransmission mechanism. 
       
    \subsubsection{Online synchronization} 
    Intermittently offline systems keep track of the current state of the secure device by recording the offline balance in a separate ledger on the online CBDC system called the \emph{offline ledger}~\cite{bis2023b}. Further, they can retrieve transaction metadata from secure devices that could be used to detect anomalies and potential instances of fraud~\cite{bis2023b}. As such, and depending on the compliance requirements of the system, some or all of the following actions take place during online synchronization:
    {\em (i)} the balance history of the device is shared with the offline ledger, 
    {\em (ii)} the transaction log, locally stored in the device, is shared with the online system to enable more thorough compliance checks, and 
    {\em (iii)} the risk parameters (\emph{e.g.,} balance, turnover limits) on the card are reset and -- if necessary -- updated to new values.

\section{System Design}\label{sec:details}
This section outlines the system’s design. It begins with an overview of the system requirements, followed by a detailed description of the main communication protocols: mutual authentication, offline payment, and bank operations.

\subsection{System requirements}
    We identify the following requirements for the proposed offline CBDC system, categorized into {\em Security} (S), {\em Privacy} (P), and {\em Compliance} (C):
   
    \begin{enumerate}
        \item[S1.] \emph{Conservation of money supply and double spending:} The system must prevent the creation of new money. Only authorized entities (\emph{e.g.,} the central bank) should be able to introduce new currency into circulation. Similarly, the system must ensure that funds cannot be accidentally or maliciously destroyed. As a by-product, users must not be able to spend the same funds more than once. 
        \item[S2.] \emph{Transaction atomicity:} All transactions must be executed atomically and consistently. Partial or incomplete transactions should be prevented as they may introduce ambiguous states or lead to the loss of funds. 

        \item[P1.] \emph{Transaction confidentiality:} The FIs should not be able to gather details of individual offline transactions, including the counterparty's identity, the transaction amount/time, and other data associated with the transaction (\emph{e.g.,} type of goods purchased, etc).
        \item[P2.] \emph{Anonymous withdrawals:} The FIs should not be able to determine the identity of users withdrawing funds.
        \item[P3.] \emph{Transaction unlinkability:} The FIs should not be able to link multiple transactions to the same user or determine their transaction patterns.

        \item[C1.] \emph{User verification:} The system must incorporate KYC procedures to verify users before allowing them to participate in the offline CBDC network.
        \item[C2.] \emph{Balance \& transaction tracking:} The FIs must be able to access the offline balance stored in the secure device, or predicates derived thereof, at regular points in time. In case of stricter regulations, it must also be able to access transaction metadata. 
        \item[C3.] \emph{Thresholds:} The secure device should enforce specific predetermined limits on the maximum balance held in it, as well as per-transaction and cumulative transaction amount limits. Those limits should be able to be
        modified by the central bank periodically so as to keep in pace with changing regulatory policies.
    \end{enumerate}
As described later, a benefit of the proposed implementation is its customizability that allows to balance AML/CFT regulation and privacy for the various requirements above. 

\subsection{Mutual authentication}\label{sec:ma_detailed}
\begin{figure}
    \centering
    \includegraphics[width=0.5\textwidth]{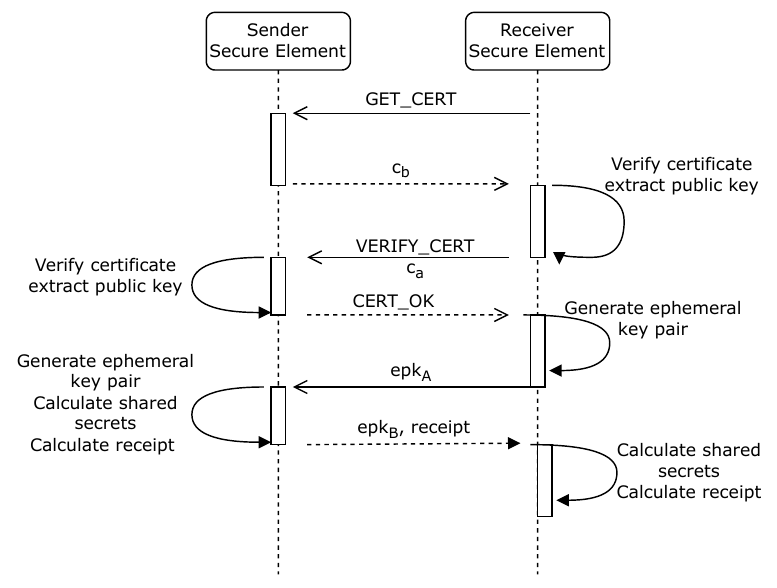}
    \caption{The mutual authentication protocol}
    \label{fig:ma}
\end{figure}

The mutual authentication protocol~\cite{scp11,nist2018} is used as a subroutine by the other system protocols to create a secure and authenticated communication channel that ensures message confidentiality and integrity. Fig.~\ref{fig:ma} presents a high-level description of the protocol between two secure devices. It does not include the wallet applications, which act only as message forwarders. The same protocol also secures the communication between a secure device and its wallet application, and between wallet applications themselves.

\subsubsection{Certificate exchange}
Both secure devices involved in the protocol, named $d_A$ and $d_B$ for Alice and Bob respectively, have a static pair of long-term keys $(sk_{i},\, pk_{i})$~\cite{nist2018} and a certificate $c_{i}$, $i\in\{A,B\}$.
First, $d_A$ requests $d_B$'s certificate $c_B$, verifies it using the public key of the certificate authority, and extracts $pk_B$. Similarly, $d_A$ responds with its own certificate $c_A$, which $d_B$ processes in the same way.
This exchange allows both parties to authenticate to each other and verify that they are communicating with authorized devices within the jurisdictional ``CBDC perimeter''. It also enables the SE to determine which operations of its API can be invoked by the counterparty. 

\subsubsection{Key establishment}
During this stage, devices negotiate and derive a shared secret which is used to create session-specific symmetric cryptography keys that encrypt and verify all subsequent communication.
First, $d_A$ generates an ephemeral key pair $(esk_A,\, epk_A)$ that is valid only for the length of each session and is used to ensure \emph{forward secrecy} (\emph{i.e.,} the confidentiality and integrity of any intercepted past communications is ensured in case the device's static keys are exposed at a later point in time).
Next, $d_A$ sends $epk_A$ to $d_B$, which in response generates $(esk_B,\, epk_B)$ and computes $Z_s =  DH(sk_B,\, pk_A)$, $Z_e = DH(esk_B,\, epk_A)$, and $Z = Z_s || Z_e$, where $DH$ is a Diffie-Hellman cryptographic primitive used to generate the static and ephemeral shared secrets $Z_s$, $Z_e$.
Then, a Key Derivation Function (KDF) is applied to $Z$ to derive three symmetric keys: an encryption key $S_{\mathrm{ENC}}$ for ensuring data confidentiality used for encrypting exchanged data, a Message Authentication Code (MAC) key $S_{\mathrm{MAC}}$ used to produce MACs~\cite{Goldreich2004} of the data to ensure data integrity and authenticity, and a receipt key $S_\mathrm{R}$ for verifying successful key establishment.
Lastly, $d_B$ responds with $epk_B$ and a cryptographic receipt created using $S_\mathrm{R}$.

Device $d_A$ repeats the above procedure to reproduce the symmetric keys by calculating $Z_s =  DH(sk_A,\, pk_B)$ and $Z_e = DH(esk_A,\, epk_B)$, and verifies the receipt.
At this point, a secure channel has been established and all subsequent communication is protected through encryption and MACs. 

\subsection{Offline payment protocol}\label{sec:vex_prot}
\begin{figure}
    \centering
    \includegraphics[width=0.5\textwidth]{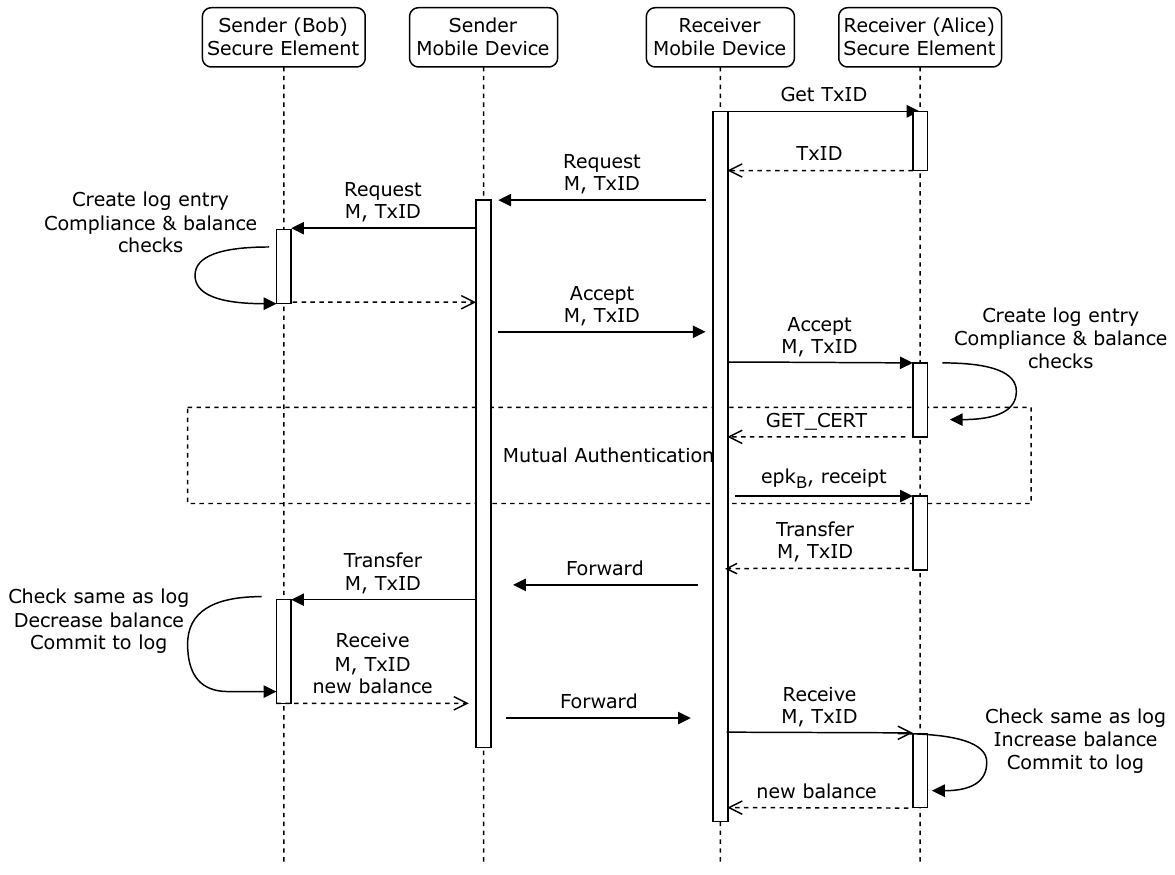}
    \caption{The offline payment protocol}
    \label{fig:payment_details}
\end{figure}

The offline payment protocol enables secure peer-to-peer transfers of funds between offline devices. The protocol, depicted in Fig.~\ref{fig:payment_details}, consists of two stages: payment initiation and value exchange. 

\subsubsection{Payment initiation} The payment workflow starts with Alice authenticating to her $d_{A}$ (via PIN, biometrics, etc. protected by the secure channel already established between $d_{A}$ and $w_{A}$) and inputting the desired amount $\mathcal{M}$ into $w_A$. The wallet requests the creation of a random transaction identifier $\mathrm{TxID}$ from the SE, which also stores the identifier internally. Then, $w_{A}$ constructs a \texttt{<Request>} $(\mathcal{M},\, \mathrm{TxID})$ and transmits it to the sender's wallet application $w_{B}$. 

Upon receipt, $w_B$ presents the payment request to Bob for approval. In case of rejection, an appropriate message is returned to Alice and the protocol terminates.
In case of acceptance, the payment request is forwarded to $d_{B}$ which performs checks to verify that: \emph{(i)} the offline balance is sufficient; \emph{(ii)} the requested amount does not exceed single-transaction limits; and \emph{(iii)} cumulative transaction limits are not exceeded. 
If any check fails, a rejection message is returned. Otherwise, $d_{B}$ creates a pending transaction entry in its secure log and $w_{B}$ constructs an \texttt{<Accept>} $(\mathcal{M},\, \mathrm{TxID})$ message. Since at any given time only one pending transaction can exist, $d_{B}$ will overwrite the latest transaction if it is still marked as pending. 
Then, $w_B$ transmits the acceptance message back to $w_A$, which checks it and forwards it to $d_{A}$. The latter performs similar compliance checks, and if successful, a corresponding pending transaction entry is created in its secure log. If not, a rejection message is returned to $w_{B}$.

At this point, both secure elements have committed to the transaction details but have not yet modified their balance counters. To proceed, $d_{A}$ initiates a request for the establishment of an end-to-end session with $d_{B}$. If one of the two devices cannot produce the appropriate certificate, the protocol aborts, otherwise it proceeds according to Section~\ref{sec:ma_detailed}, with $w_{A}$ and $w_{B}$ acting as forwarders facilitating the communication. When completed, all subsequent communication between $d_A$ and $d_B$ is encrypted and authenticated. 

\subsubsection{Value exchange}
After establishment of the end-to-end communication channel, $d_{A}$ issues a \texttt{<Transfer>} command containing $(\mathcal{M},\, \mathrm{TxID})$, to be executed by $d_{B}$. Upon receiving this command, the sender's SE checks that the payment details match the pending transaction in its log and will atomically: \emph{(i)} decrement its balance and \emph{(ii)} update the transaction log with $pk_{A}$ as the counterparty ID and set the transaction status to ``completed''.
As a last step, it responds with a \texttt{<Receive>} command containing $(\mathcal{M},\, \mathrm{TxID})$ and also reports the new balance to $w_{B}$.
The response of $d_{A}$ to the command is similar: it first verifies that payment details match its pending transaction and then atomically: \emph{(i)} increments its balance and \emph{(ii)} updates the transaction log with $pk_{B}$ and sets the status to ``completed''. Finally, it reports its new balance back to $w_{A}$.

\subsubsection{Offline payment with one user wallet}
Notably, it is not necessary for both users to have a wallet application~\cite{BoN2025};
the system can serve transactions when one user has only a secure device and the other has a ``trusted'' Point of Sale (POS) terminal (such as those available in retail stores today) as long as it contains an SE to execute the protocol. In this case, the POS assumes the roles of $d_{A}$, $w_{A}$, and $w_{B}$.

\subsection{Retransmission mechanism}
The retransmission mechanism allows devices to repeat the last transaction if the communication was interrupted. Whether the mechanism is used depends on which point the execution of the protocol fails.
Specifically, if it is interrupted at any point during the payment initialization phase, then the devices restart the payment by overwriting the pending state without the need for retransmission. Similarly, if it is interrupted before the \texttt{<Transfer>} command reaches the sender, then the devices start over. 
On the contrary, if communication is lost after $d_{B}$ decrements its balance but before $d_{A}$ increments its own, then Alice and Bob can immediately engage in the retransmission protocol. The wallet application of Alice issues a \texttt{<Retransmit>} command to $d_A$, containing the amount and $\mathrm{TxID}$ of the failed transaction. Her secure device checks the most recent entry in its log to verify that \emph{(i)} the transaction details match and \emph{(ii)} the status is ``pending''.
If both conditions hold, then it proceeds with establishing a new secure session with $d_B$ and repeating the value exchange phase by issuing a \texttt{<Transfer>} command. Upon receipt, $d_B$ checks its most recent entry in its log to verify that \emph{(i)} the transaction details match and \emph{(ii)} the status is ``completed'' and issues a \texttt{<Receive>} command without decreasing its balance counter. Upon receipt, $d_A$ completes the phase as described in the previous section.

\subsection{Bank operations}
When Alice connects to the FI terminal, first a secure session is established between $d_{A}$ and the terminal, then she authenticates herself to $d_{A}$, and finally the online synchronization protocol is automatically run followed either by a  withdraw or a deposit based on the input from Alice.

\subsubsection{Online synchronization}\label{sec:online_sync} The protocol begins with the FI terminal issuing a \texttt{<Synchronize>} command to the secure device, which responds with its current balance and either only the amounts of the stored transactions or with all the metadata (\emph{i.e.,} transaction ID, counterparty ID, amount, and transaction status).
The FI will then replay these transactions, applying them to the last recorded balance in the offline ledger before the synchronization and if the synchronization is successful -- the result matches the current offline balance -- the device is instructed to clear its log and reset its threshold limits. Otherwise, the card is blocked and the protocol aborts.

\subsubsection{Withdraw \& deposit}
In case of a withdrawal, the FI terminal decreases the online balance of the user (if sufficient to provide the requested amount) and issues a \texttt{<Withdraw>} command containing the amount. The secure device checks if the amount does not exceed the balance limit and proceeds accordingly. In case of a rejection, the terminal is informed to revert the online balance.
When depositing, the terminal issues a \texttt{<Deposit>} command containing the requested amount. The secure device checks if there is enough balance on the device and either accepts the request and returns the new balance or rejects it. Upon receiving a positive response, the FI will increase the user's balance.

\begin{figure}
    \centering
    \includegraphics[width=0.7\linewidth]{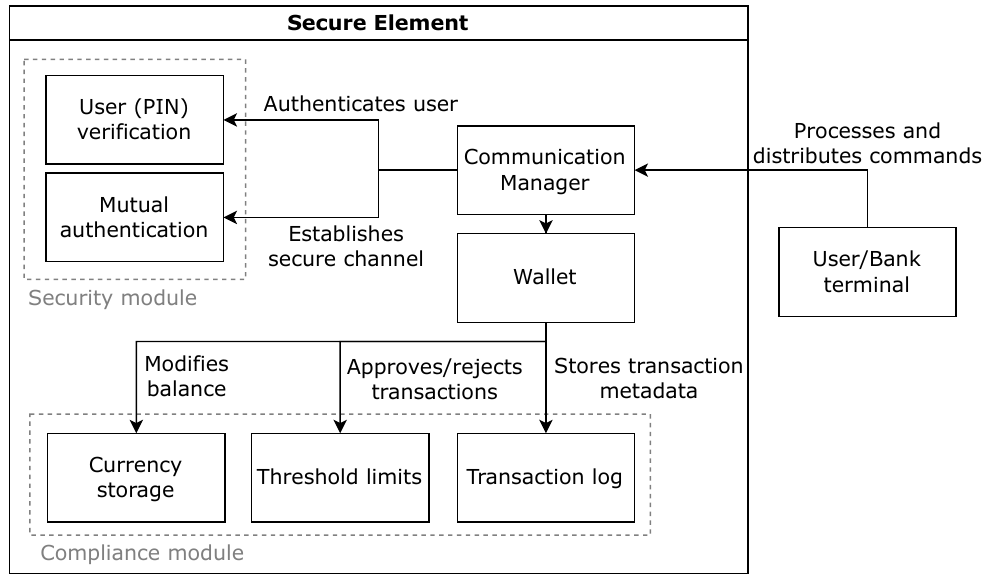}
    \caption{Module-level architecture of the secure device}
    \label{fig:modules}
\end{figure}

\section{Prototype Evaluation}\label{sec:eval}
In Fig.~\ref{fig:modules}, a module-level diagram of the system is provided.
The communication manager routes commands to system modules. The security module handles user authentication and secure session setup. Finally, the wallet communicates with the compliance module to manage the balance on the card, enforce regulatory limits, and keep track of the transactions.

Implementation-wise, the prototype uses Elliptic Curve Cryptography with keys defined over the secp256r1 curve. AES-256 is used for session keys, with a KDF based on the ANSI X9.63 standard. SHA-256 is used for hashing, and certificates follow the format specified in~\cite{scp11}.
Experiments use two ACOSJ 95K smart cards (Java Card 3.0.4, Global Platform 2.2.1) as secure devices and simulated user and FI terminals on a computer with an Intel i7-10750H processor and 32\,GBs of RAM running Windows 11.

We consider three compliance cases. The first (``compliance-free'') does not have any online synchronization constraints, while the second and third implement balance and transaction tracking, respectively, by following the procedure described in Section~\ref{sec:online_sync}.
To further study the impact of the mutual authentication protocol on the performance of the system, we implement two variants of the prototype: \emph{V.1} and \emph{V.2}. The first executes the mutual authentication protocol as described in Section~\ref{sec:ma_detailed}, whereas the second explores a trade-off between security and performance.
Specifically, instead of using ephemeral key pairs, the initiating party (\emph{i.e.,} the receiving SE or a terminal) generates and sends a nonce value~\cite{nist2018}.
Under this scheme, the shared secret $Z$ depends only on the static key pairs and remains the same across sessions, while the nonce is used as a source of randomness in the KDF to ensure unique session keys.
Therefore, if the static keys are compromised, forward secrecy may no longer be guaranteed~\cite{nist2018}. 

\begin{figure}
\centering
\includegraphics[width=0.5\linewidth]{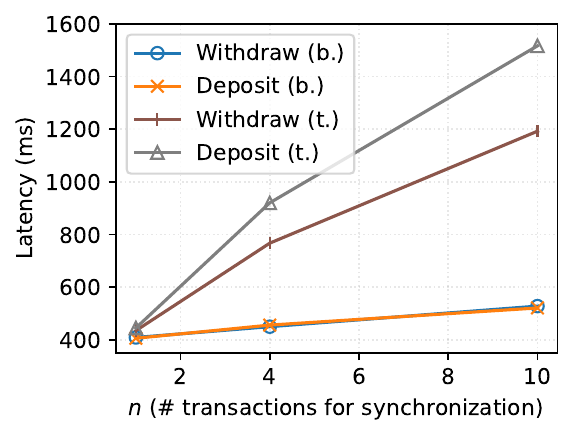}
\caption{Latency of the \emph{synchronize} sub-operation for $V.1$ for various $n$}
\label{fig:log_size}
\end{figure}

\begin{table*} 
    \centering
    \renewcommand{\arraystretch}{1.2}    
    \caption{End-to-end latencies of the basic operations for variants $V.1$ and $V.2$}
    \hspace*{-0.35cm}\begin{tabular}{lcccccccccc}
    \toprule
    & \multicolumn{5}{c}{\bfseries V.1} & \multicolumn{5}{c}{\bfseries V.2} \\
    \cmidrule(r){2-6} \cmidrule(r){7-11}
    & \multicolumn{2}{c}{\bfseries Withdraw} & \multicolumn{2}{c}{\bfseries Deposit} & \multirow{2}{*}{\bfseries Payment (ms)} & \multicolumn{2}{c}{\bfseries Withdraw} & \multicolumn{2}{c}{\bfseries Deposit} & \multirow{2}{*}{\bfseries Payment (ms)}\\
    \cmidrule(r){2-3} \cmidrule(r){4-5} \cmidrule(r){7-8} \cmidrule(r){9-10}
    \bfseries Log size & \bfseries Bal. (ms) & \bfseries Trans. (ms) & \bfseries  Bal. (ms) & \bfseries  Trans. (ms)&& \bfseries  Bal. (ms) & \bfseries  Trans. (ms) & \bfseries Bal. (ms) & \bfseries  Trans. (ms)& \\
    \midrule
    $n=0$ & \multicolumn{2}{c}{2165} & \multicolumn{2}{c}{2114} & 4791 & \multicolumn{2}{c}{1738} & \multicolumn{2}{c}{1673} & 3949\\
    $n=1$ & 2488& 2526& 2528& 2572& 4813 & 2080& 2105 & 2083& 2114& 3961\\
    $n=4$ & 2556& 3007& 2557& 3014& 4792 & 2123& 2559& 2135& 2606& 3954\\
    $n=10$ & 2612& 3587& 2637& 3638& 4788 & 2189& 3195& 2211& 3186& 3945\\
    \bottomrule
    \end{tabular}
    \label{tab:overview_combined}
    \vspace{-0.5em}
\end{table*}

Table~\ref{tab:overview_combined} features an overview of the average latencies over five iterations for the main operations of the system for variants~$V.1$ and $V.2$. We conduct experiments for varying values of $n$, where $n$ is the number of transactions to be synchronized with the FI. We note that for $n=0$, we effectively have the compliance-free case.
We observe that $V.2$ mirrors $V.1$ in behavior but with lower latencies overall, owing to the reduced impact of the alternative mutual authentication scheme. Specifically, withdrawals and deposits exhibit similar latency trends, increasing with $n$, while payment latency remains constant at around 4.8~seconds for $V.1$ and 3.9~seconds for $V.2$, irrespective of $n$.
Between the different types of compliance, transaction tracking is more expensive, reaching 3.6~seconds and 3.2~seconds, due to the larger amount of data to be synchronized.
For both variants, payment latency remains comparable with modern payment systems.
\begin{table} 
    \centering
    \renewcommand{\arraystretch}{1.2}
    \caption{Detailed latencies for the sub-operations of the protocols ($n=10$)}
    \hspace*{-0.35cm}\begin{tabular}{llcccc}
    \toprule
    && \multicolumn{2}{c}{\bfseries V.1} & \multicolumn{2}{c}{\bfseries V.2}\\
    \cmidrule(r){3-4} \cmidrule(r){5-6}
    & \bfseries Operation & \bfseries Bal. (ms) & \bfseries Trans. (ms) & \bfseries Bal. (ms) & \bfseries Trans. (ms)\\
    \midrule
    \multirow{3}{*}{\rotatebox[origin=c]{90}{Withdr.}} & Mut. auth. & 1652 & 1656 & 1231& 1231\\
    & Synch. & 528 & 1192 & 526 & 1214\\
    & Withdr. & 300 & 296 & 302 & 300\\
    \midrule
    \multirow{3}{*}{\rotatebox[origin=c]{90}{Deposit}} & Mut. auth.  & 1690 & 1696 & 1245 & 1246\\
    & Synch. & 521 & 1516  & 533 & 1512\\
    & Dep. & 298 & 295 & 300 & 300\\
    \midrule
    \multirow{3}{*}[-0.30cm]{\rotatebox[origin=c]{90}{\parbox[t]{0mm}{Offline \\trans.}}} & Init. & \multicolumn{2}{c}{495} & \multicolumn{2}{c}{492} \\
    & Mut. auth. & \multicolumn{2}{c}{3491} & \multicolumn{2}{c}{2626} \\
    & Val. exch. & \multicolumn{2}{c}{673} & \multicolumn{2}{c}{692} \\
    \bottomrule
    \end{tabular}
    \label{tab:detailed}
\end{table}

Table~\ref{tab:detailed} lists the average latencies for the sub-operations of withdrawal, deposit, and payment for $n=10$.
Here, we observe that the most expensive operation is the establishment of a secure channel between the participants, reaching 3.5~seconds (72.9\,\% of the total latency) for payments under the $V.1$ variant. This indicates a clear bottleneck of the system and an avenue for future improvements. The next most expensive operation is the synchronization in the transaction tracking case taking as much as 1.5 seconds.

In Fig.~\ref{fig:log_size}, we examine the impact of $n$ in the latency of the synchronize sub-operation for 1, 4, and 10 transactions under balance and transaction tracking for $V.1$.
Latency increases with the number of transactions to be synchronized, and this increase is more prominent during transaction tracking.

\section{Privacy Preservation Through ZKPs}\label{sec:zkps}
In this section, we discuss potential enhancements to the system through the combination of ZKPs and Verifiable Credentials (VCs)~\cite{Sporny2022} -- secure and verifiable digital representations of documents attesting identity attributes (\emph{e.g.,} passports, diplomas), sometimes also called attestations of attributes -- to minimize the exposure of users' sensitive information.

\subsubsection{Verifiable transactions}
The offline payment protocol (see Sec.~\ref{sec:vex_prot}) can be extended by embedding conditions related to the users' identities during the payment initiation phase.
More specifically, an extra {\tt conditions} field is added to the payment request (\emph{i.e.,} $(\mathcal{M},\, \mathrm{TxID},\, {\tt conditions})$) through which the receiver can request a ZKP on an attribute of the sender (\emph{e.g.,} age) as an extra condition to accept the payment. If the payer satisfies the condition(s) specified by the receiver, they append a ZKP $\pi_1$ to their response: \texttt{<Accept>} $(\mathcal{M},\, \mathrm{TxID}, \, \pi_1)$.

More specifically, the payer has to provide a ZKP attesting to the following, as also depicted in Alg.~\ref{alg:zkp1}: \emph{(i)} it posses a valid VC signed by the certificate authority;
\emph{(ii)} it has control over the public key included in the VC; and
\emph{(iii)} the VC contains an attribute whose value is above a given threshold.
We note that the proof is verified by $w_{A}$ and not by $d_{A}$, since these kind of conditions are not directly related to the security of the payment system, and thus, not a concern to the SE. For example, if a store sells alcohol to an underage person, it does not directly affect the integrity of the payment system. Therefore, the wallet's software can more flexibly accommodate changes in regulation or complex logic as compared to the applet running on the SE. Moreover, the corresponding VCs and cryptographic keys may not be directly accessible to the SE, as some regulated wallets require a separate secure device for managing the private keys corresponding to the VCs.  

\begin{algorithm}
\caption{Zero-Knowledge Proof $\pi_1$}
\label{alg:zkp1}
\begin{algorithmic}
    \STATE \KwPubIn{threshold $\tau$, attribute $k$, CA public key $pk_{CA}$, transaction ID $\mathrm{TxID}$, timestamp $t_{now}$}
    \STATE \KwPrivIn{Verifiable credential VC, public key $pk$, signature $s$}
\end{algorithmic}
\begin{algorithmic}[1]
    \STATE \KwAssert{isSignatureValid(VC, VC.sig, $pk_{CA}$)}\;
    \STATE \KwAssert{$VC.expiry > t_{now}$}\;
    \STATE \KwAssert{$VC.pk == pk$}\;
    \STATE \KwAssert{isSignatureValid($\mathrm{TxID}$, $s$, $pk$)}\;
    \STATE \KwAssert{$VC.k \geq \tau$}\;
\end{algorithmic}
\end{algorithm}

\subsubsection{Anonymous withdrawals}
The mutual authentication protocol with the FI during CBDC withdrawal can be modified to allow for privacy-preserving withdrawals, where the user exchanges physical cash for CBDCs at an FI terminal.
Specifically, the secure device does not make use of its static key but instead uses only an ephemeral one ($epk$), thus protecting the user's identity from the FI. On the other hand, the FI uses only its static key. Therefore, the FI is identified through its public key, whereas the wallet application provides a ZKP instead, which attests to the following: 
\emph{(i)} it possesses a valid VC signed by the certificate authority; \emph{(ii)} it has control over the public key included in the VC by using it to produce a signature on the ephemeral key. These operations correspond to lines 1-4 of Alg.~\ref{alg:zkp1}, with $epk$ replacing $\mathrm{TxID}$.

\section{Security Analysis}\label{sec:sec_analysis}
In this section, we conduct a security analysis of the proposed system under the following assumptions:
\emph{(i)} the FI is honest but ``curious.'' It will not deviate from the protocol, but remains interested in collecting information about the participants;
\emph{(ii)} users can be malicious and may attempt to double spend or compromise their devices;
\emph{(iii)} the cryptographic keys of the FI are securely stored and cannot be compromised;
\emph{(iv)} the cryptographic primitives used by the protocols are safe against compromise. 
We also discuss the system's flexibility to accommodate trade-offs between privacy and transparency.

\subsubsection{Adversary}
We assume a powerful adversary with knowledge of the protocol that is capable of observing and manipulating all communication channels by intercepting, modifying, dropping, or delaying messages between legitimate parties.
They can also initiate new sessions with legitimate participants.
The adversary is computationally bounded and cannot break established cryptographic primitives.

\subsubsection{Conservation of money supply and double spending}
An adversary can illegitimately create new money only if they manage to convince their secure device to accept either a withdrawal confirmation or a \texttt{<Receive>} message.
First, the adversary would have to impersonate the FI (by presenting an FI terminal certificate during authenticated session establishment) and participate in the withdrawal protocol with the secure device. Since by Assumption~\emph{(iii)} the secret key of the FI cannot be compromised, the adversary cannot prove control over an FI terminal certificate and therefore cannot use this route.
Similarly to send a \texttt{<Receive>} message, the adversary would have to possess a secure device key to impersonate one. Assuming that SEs cannot be compromised (a possibility examined later in this section), this would be impossible.

Further to this, when a receiving secure device creates a payment request, it generates a random transaction ID and keeps it in its memory as the current ID.
At any subsequent point in time and until the transaction is finalized (\emph{i.e.,} the balance counter update is committed), the device will accept a \texttt{<Receive>} message only if it contains that particular ID.
Therefore, intercepted \texttt{<Receive>} messages from previous transactions cannot be replayed and will be rejected. An immediate corollary is that an adversary cannot double spend. 

An adversary participating as the receiver in a transaction would be able to reduce the money supply by dropping the \texttt{<Receive>} message. This would cause the sender to reduce their balance, but the receiver's balance would not increase effectively forfeiting the funds. However, such an attack would require the adversary convincing other users to transact with them (evidently in exchange for a good or service, thus incurring a cost to the adversary) or illegally obtaining a large amount of secure devices and circumventing the user authentication step.
During bank transactions new money cannot be created but it can be possibly destroyed. In that case, the offender will be detected at the next synchronization, this being an intermittently offline CBDC, as the online balance accounting will not match the secure device's reported values.

\subsubsection{Transaction atomicity}
If the receiving party of a transaction is not adversarial, then the retransmission protocol ensures that interrupted transactions that result in an incomplete state can be repeated, ensuring a consistent state between the two participants and no loss of funds.

\subsubsection{Compromised mobile device}
Since the phone of the user is considered untrusted, an adversary could extract its secret key and certificate and impersonate it, thus forcing the SE to accept commands from a fake user terminal. Nevertheless, this does not pose a security risk, since the SE will only accept payments signed by another SE (which we assume its keys cannot be extracted). Also, the withdrawal protocol will only accept commands by the FI (whose identity is established through its certificate during mutual authentication).

\subsubsection{Compromised secure device}
If an adversary manages to compromise a secure device and extract its keys~\cite{roche2024}, then they are able to create new money by participating in the offline payment protocol and forging \texttt{<Receive>} messages. A mitigating factor in this situation would be the limits imposed by the receivers' legitimate devices, which would refuse to accept funds after a certain amount, restricting how much an adversary could double-spend per interaction.

\subsubsection{On the privacy and transparency conundrum}
The proposed system allows for the realization of different trade-off levels between
privacy and transparency~\cite{Michalopoulos2025,pocher2022}. Specifically we distinguish between the following scenarios: If the CBDC system has provisions for anonymous withdrawals and no transaction and balance tracking are required, then all privacy properties (P1-P3) are satisfied by the prototype giving rise to its most private version that can ensure the anonymity of the user toward the FI. If balance reporting is enforced, then transaction confidentiality can no longer be guaranteed, leading to a moderately private version (P2, P3). Finally, when both forms of monitoring (balance and transaction) are implemented, the user has a low degree of privacy (in the case of anonymous withdrawals) (P2), or no privacy at all.

\section{Conclusion and Future Directions}\label{sec:conclusion}
As Central Banks look to redefine the very essence of cash with CBDCs, this paper presented a proof-of-concept implementation for offline digital currency transactions that balances the competing requirements between privacy and regulatory compliance. 
Its findings indicate that the proposed implementation safeguards payment integrity and that it is suitable for real-life daily use. Several challenges remain to be addressed in future work.
First, a formal verification of the system's protocols could provide stronger guarantees about its resistance to various attacks.
Next, the incorporation of revocation lists could allow for more effective control of compromised devices and strengthen accountability. 
Finally, while the paper hints on the interplay between CBDCs and VCs, a stronger integration of CBDCs with emerging digital identity systems (such as the European digital identity wallets) could function as a deterrent to illicit activities (\emph{e.g.,} implement all-or-nothing transferability by amplifying the consequences of device sharing for circumventing regulatory limits) and enable richer functionality for both systems~\cite{Gross2021}.
The offline CBDC landscape may be at its infancy, but this work demonstrates that viable combinations of hardware and software can address some of the key challenges of offline CBDCs.

\bibliographystyle{IEEEtran}

\bibliography{arxiv}

\end{document}